# A multipurpose action for learning/teaching process: The Pigelleto's Summer School of Physics

*Vera Montalbano[1], Emilio Mariotti[1]*
[1]Department of Physical Sciences, Earth And Environment, University of Siena, Italy.

*Abstract:* Since 2006, forty students from high school are selected to attend a full immersion summer school of physics in the Pigelleto Natural Reserve, on the south east side of Mount Amiata in the province of Siena. Topics are chosen so that students are involved in activities rarely pursued in high school, aspects and relationship with society are underlined and discussed. Our purpose is offering to really motivated students an opportunity of testing the scientific method, the laboratory experience in a stimulating context, by deepening an interesting and relevant topic in order to orienting them towards physics. Students are encouraged in cooperating in small groups in order to present and share the achieved results. Starting from the third edition of the school, the school became a training opportunity for younger teachers which are involved in programming and realization of selected activities. The laboratory activities with students are usually supervised by a young and an expert teacher in order to fix the correct methodology. Recently, young teachers enrolled in a master in Physics Education tested in the summer school some activity designed in their courses.

## BACKGROUND AND FRAMEWORK

In the past decades, an impressive decreasing of interest of young people in pursuing scientific careers was observed almost everywhere in the world (Czujko, 2002; Convert, 2005; National Science Board, 2007; Mulvey & Nicholson, 2011). In the last years, many Italian Universities are involved in a large national project in order to enhance the interest of high school students towards scientific degrees (Sassi, Chiefari, Lombardi, & Testa, 2012; Montalbano, 2012).

In this context, we started to organize a full immersion summer school of physics in the Pigelleto's Natural Reserve, on the south east side of Mount Amiata in the province of Siena. In the beginning, our purpose was orienting talented students by offering an opportunity of testing the scientific method through experience in physics laboratory in a stimulating context. The students (age 16-17) are selected by their teachers in the network of schools involved in the National Plan for Science Degree. We usually propose participate lessons in which the necessary background for the following activities in laboratory is given. Small groups of students from different schools and classes are engaged in inquiry-based laboratories where they are encouraged to take an active role (Bonwell & Eison, 1991).

### National Plan for Science Degrees

Italy is active since 2005 with a large plan to contrast decreasing in enrollment in basic sciences, i.e. Mathematics, Physics, Chemistry and Science of Materials.

This action originated from a collaboration of the Ministry of Education and Scientific Research, the National Conference of Deans of Science and Technology and Confindustria, the main organization representing Italian manufacturing and services companies.

The plan is focused on the following main objectives:

- improving knowledge and awareness of science in secondary school, offering students in the last three years of school to participate in stimulating and engaging curricular and extracurricular activities in laboratory;

- starting a process of professional development of science teachers in service in the secondary school from joint work between School and University for the design, implementation, documentation and evaluation of the laboratories mentioned above;

- promote alignment and optimization of training from University and School for the working world;

- the necessity to revise the content and methods of teaching and learning of science in all grades of school, taking into account the new national guidelines for first and second cycle contained in the recent Italian reform of the educational system.

In order to achieve the above purposes, strategy and methodologies pursued and realized are:

- orientation is an active action that the student achieves by perform meaningful activities that allow to compare problems, issues and ideas of science;

- designing the training of teachers by involving them in solving concrete problems, developing design and implementation of educational activities and through comparison with peers and experts;

- pursuing and achieving at the same time the student orientation and training for teachers through the planning and joint implementation by school teachers and university laboratories for students, thus developing relations between the school system and the University;

- consciously connecting the activities of the Plan with the innovation of curricula and teaching methods adopted in schools, and other contents and methods of teacher training (initial and in-service), for the first and second cycle.

Thus, laboratory is a method, not as a place, students become the main character of learning and joint planning by teachers and university is a mandatory step.

In this context, we started to design a full immersion summer school in physics together with some expert teachers . This small team (three from university and three from high school experience) had a parallel experience in pre-service teacher education and shared the same idea of active and inquiry forms of learning.

# THE PIGELLETO'S SUMMER SCHOOL OF PHYSICS

The school was born eight years ago and starts usually in the beginning of September. The 2013 edition is titled *The laser*, some previous editions were *Light, color, sky: how and why we see the world* (2006), *Store, convert, save, transfer, measure energy, and more…*(2007), *The achievements of modern physics* (2009), *Exploring the physics of materials* (2010), *Thousand and one energy: from sun to Fukushima* (2011).

Forty students from high school are selected to attend at a full immersion summer school of physics in the Pigelleto Natural Reserve (see fig.1), on the south east side of Mount Amiata in the province of Siena. The students (age 16-17) are selected by their teachers in a wide network of schools within the National Plan for Science Degree and come from southern Tuscany (Arezzo, Siena, Grosseto).

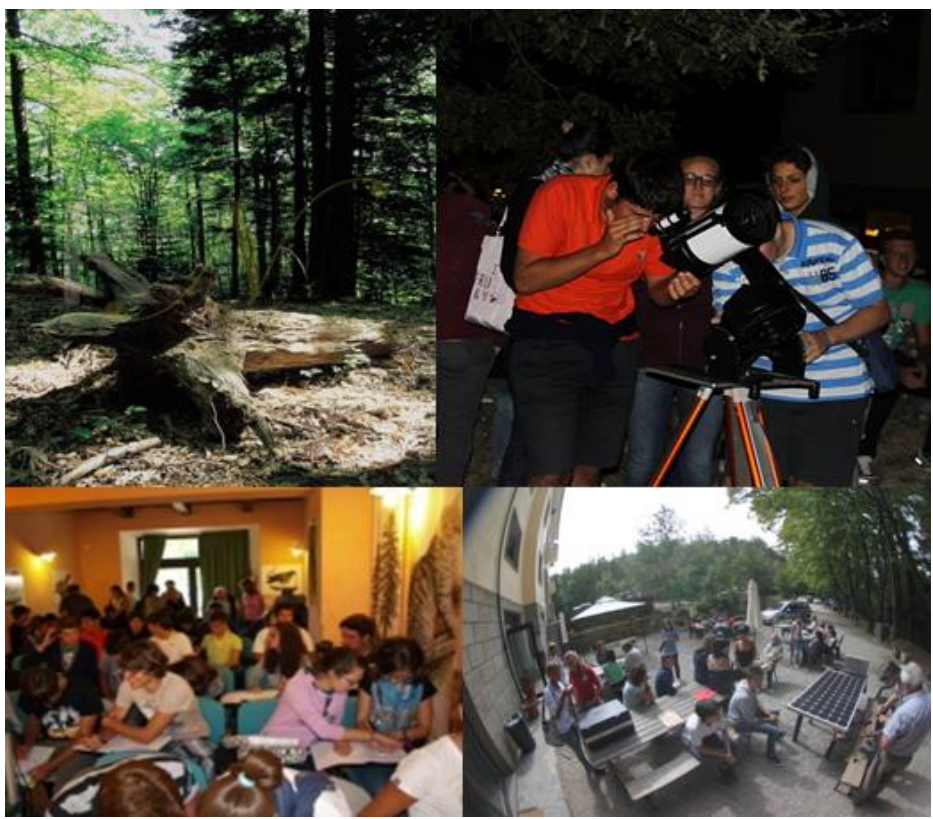

*Figure 1*. A view of woods in natural reserve, astronomical observation of the sky, a participate lesson and waiting for lab starts.

Topics are chosen in such a way that students are involved in activities rarely pursued in high school, aspects and relationship with society are underlined and discussed.

In the morning, we propose lessons in which the necessary background for the following activities in laboratory is given. In the afternoon, small groups of students from different schools and classes are engaged in laboratories where they can play an active role, which is encouraged throughout all school activities. All groups are supported by one or two teachers that are available to discuss any idea.

Usually we propose different laboratories and each group of participants must prepare a brief presentation for sharing with all other students what they have learned. After dinner, an evening of astronomical observation of the sky is usually expected. If it is cloudy, a problem solving evening is proposed.

Many educational choices are involved in designing the summer school and main target and methodologies are discussed and selected with the teachers involved in PLS:

- almost all laboratories are made with poor materials or educational devices provided by some schools;
- the groups are inhomogeneous and formed by following the teachers suggestions in order to have the best collaboration;
- the main topic is related to all activities and must be not trivial;
- almost all laboratories lead to at least one measure and the error valuation;
- in the lab, there is always at least an activity that we can call How it works.

## A full immersion experience for students

The summer school lasts four days (about 70 hours) and usually students are engaged in activities for about 32 hours. Students are housed in dormitories (6-8 beds), eating together and usually discuss about physics at any time (during breaks, at night, etc.), well beyond our initial expectations showing a deep involvement (fig.2). An important aspect is the full immersion in the beautiful natural surroundings that positively predisposes and favors concentration.

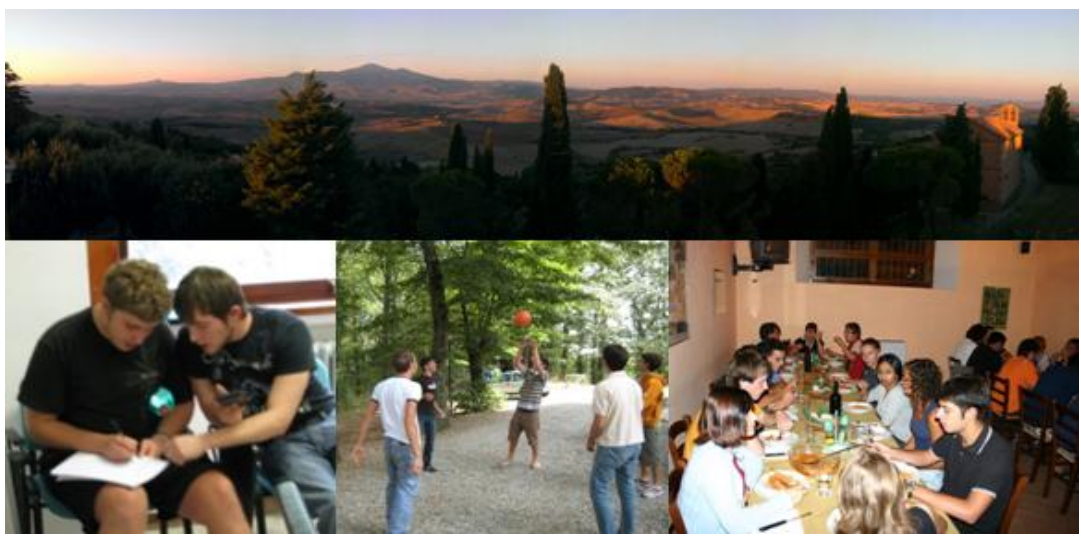

*Figure 2*. The landscape from Pienza, where the school temporary moved in 2013, with Mount Amiata on the left, students during breaks and meals.

Introductory participate lessons, inquiry-based laboratories, peers presentations for sharing knowledge are proposed for obtaining a better understanding a meaningful topic. The final step is more orienting to Science Degree and consists in one or two seminars given by young physicists (usually Ph. D. students) which can be called Insight labs, where some research in the topic of the school is presented.

In Table 1, lessons and laboratories are showed for three editions of the summer school.

Table 1

*Examples of activities framework in the summer school.*

| School's edition | *The achievements of modern physics* | *Exploring the physics of materials* | *Thousand and one energy: from sun to Fukushima* |
|---|---|---|---|
| Lessons | On the validity of a Physical theory  On concepts of classical physics disproved by modern physics  Atomic models: from classical to modern  How laser works  Radioactivity, nuclei and surrounding  On nuclear energy and energy resources | Discovering materials  Physics of polymeric materials  Terrestrial and stellar density  Materials science and biomedicine | On energy  Energy from nuclei  Energy from sky and sun  Energy from the ground  On photovoltaic panels  On fuel cells  Conductors and superconductors |
| Laboratories | How it works: a spectroscope  Measurement of Plank's constant with a LED  Photoelectric effect  We characterize a diode laser  We characterize the natural radioactivity  Measure of the speed of rotation of a star | How it works: plastic bottles  Birefringence induced by mechanical stress  Physics of Growing-Spheres  Semiconductors and photodiode.  Ferrofluid and magnetic field  Optical properties of an alkaline vapour lighted by a laser diode  Measurement of resistivity of soil  Superconductors and surroundings | How it works: a photovoltaic panel  How it works: Stirling machine, solar oven, coffee pot  Measurement of the mechanical equivalent of heat  Induction and friction  Newton's cradles and Yo-Yo  Radioactivity background vs. Weak Uranium sources |

*Focus on laboratories*

All activities in the summer school are focused on laboratory, whenever it is possible inquiry-based labs. Teachers propose experimental situations that small groups of students can investigate (fig. 3). Usually at least 8 different labs are proposed (corresponding to the maximum number of spaces which can be used for these activities), sometimes more (in this case labs are different every day).

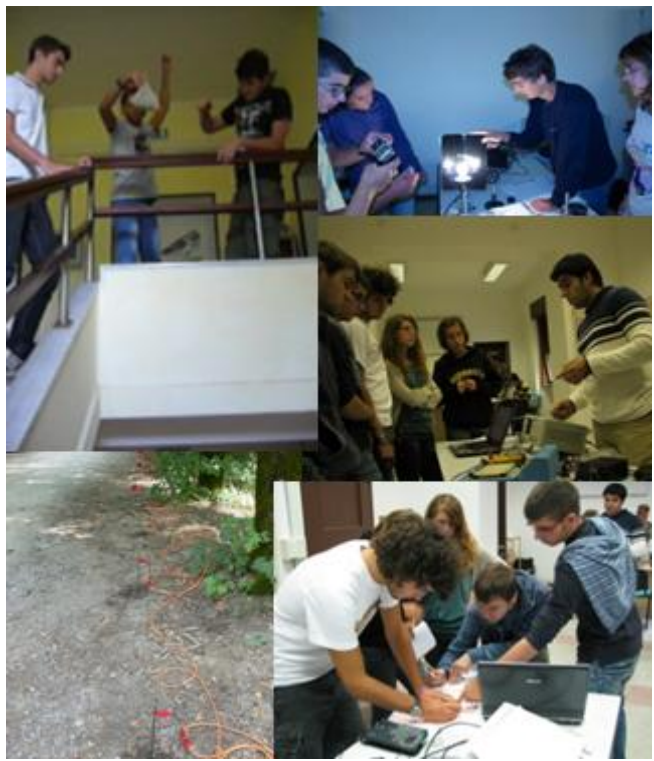

*Figure 3.* Examples of laboratory: transforming potential energy in heat, studying a light source, characterizing a diode laser, measuring resistivity of soil, analyzing and discussing data.

All labs are connected and to the main topic proposed by the school and activities are arranged in such a way that active and collaborative learning is encouraged. A puzzle or a paradox involving the concept at issue can be presented in order to struggle towards a solution, engaging students to work it out without authority's solution.

A closer analysis (Benedetti, Mariotti, Montalbano, & Porri, 2011) showed that all requests for achieving a cooperative learning (Curseo 1992, Johnson 1999) were satisfied.

## Students Communicate Physics

The oral talk in which students share what they have learned in laboratory is stressed as the central part of the school (see fig. 4). Students are not used to speak in public and fear this moment but usually the most part of them responding to the challenge. They can choose the form and the means utilized in the communication. Everyone can talk or a group can delegate a speaker, they can present the data in the blackboard or with slides. Furthermore, they can show devices or material for better explain the topic. At the end a brief discussion is performed between the communicators and the public.

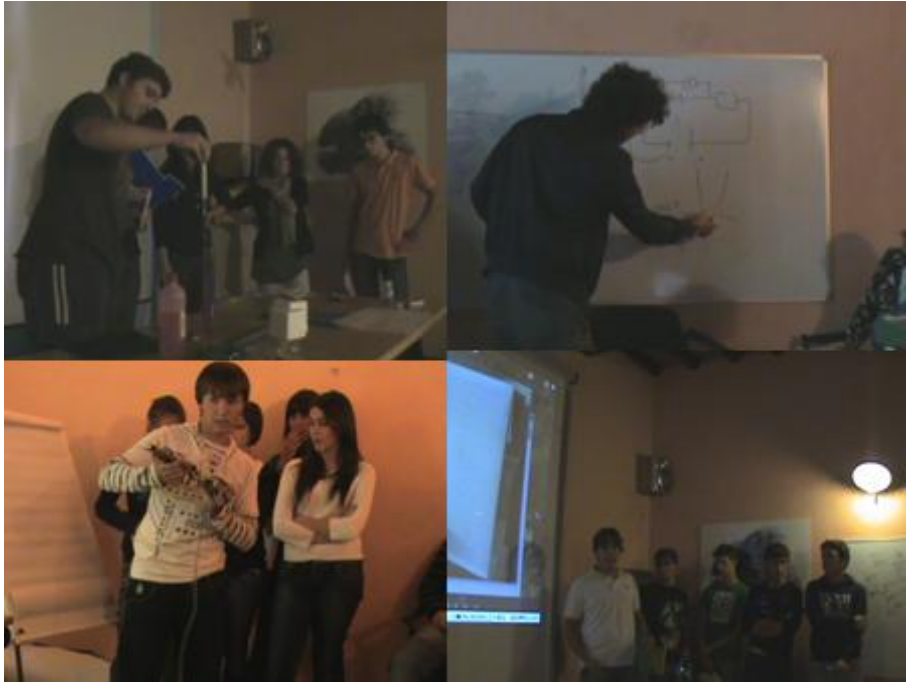

*Figure 4.* Some presentation: students can choose the form and the means utilized in the communication. Everyone can talk, a group can delegate a speaker and can present the data in the blackboard or with slides. Furthermore, devices or material can be shown for better explanation.

In the beginning the communications were proposed in order to share the results of each lab. It was a necessity because we hadn't enough time and space to perform all experiments with everyone. In the years, communication is became the central activity in which students are involved and it encourages and increases active and collaborative behavior in the school.

## An active experience for young teachers

Starting from 2007, physics teachers enrolled in Advanced School for Teaching in Secondary School of Tuscany were invited for a stage at the summer school. After the closure of the Advanced School in 2009, we still continue to invite young teachers for a full-immersion stage.

Usually there are about 4-5 expert teachers and 5-7 young teachers at the summer school. Methodologies were discussed and selected with the teachers involvement and usually an expert and a young teacher were engaged in laboratory for improving teacher practice (Montalbano et al., 2012), as shown in fig.5.

In recent years, a national master designed for qualified teachers has been organized by University of Udine. We joined the master's third edition where nineteen universities all over the country collaborated for giving courses in laboratory, often focusing on laboratories performed within the National Plan (Stefanel et al., 2012).Teachers enrolled in the master, which were following activity at University of Siena, participated actively in the 2011 summer school.

This informal training is very effective. Participants sometimes seem to be inspired by the school's laboratories and some activities entered in practice (Montalbano et al., 2012). In last edition, some young teachers became enough expert for designing and performing by themselves one or more laboratory. They are collaborating actively in the actual edition of the school and we can say that from Pigelleto's summer school a little professional community was born.

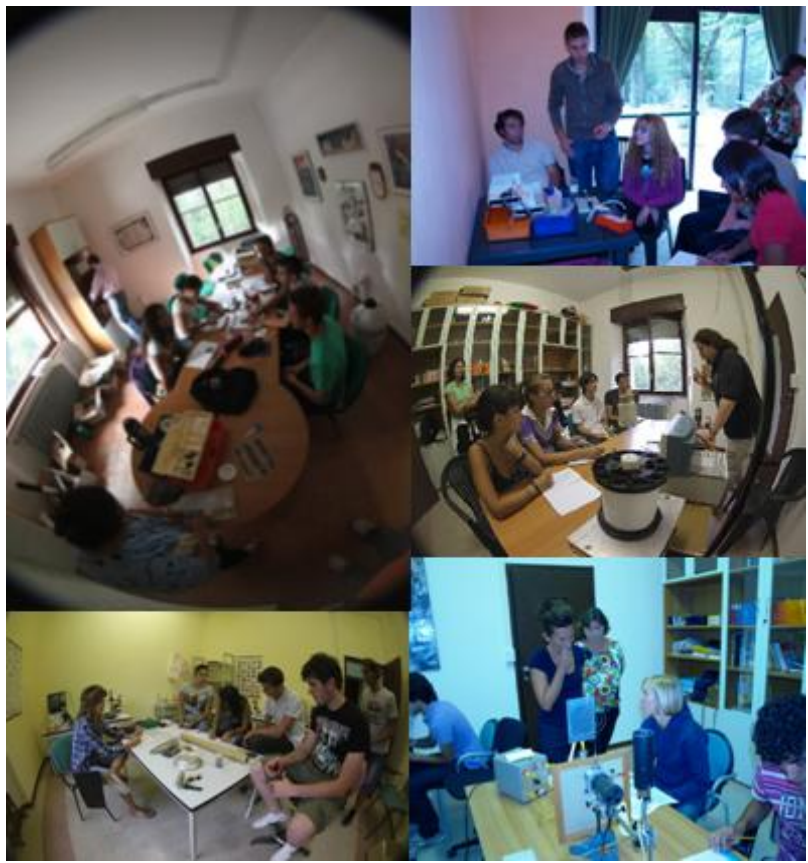

*Figure 5.* Expert and young teachers support together students in labs.

In the last academic year, a new course started in Italy in order to form teachers and it is called Formative Active Training. We invited all young teachers to attend at the summer school in 2013 but no one came.

In our opinion, the time schedule of Formative Active Training (an annual course vs. the biennial previous one, mandatory for obtaining the teacher qualification) is too compressed. Thus, teachers in training must face their final examination for qualification before the summer school, just the time in which we usually design and prepare the activities for the school's next edition.

## RESULTS AND DISCUSSION

The Pigelleto's summer school of physics remains the more effective and successful action that we perform in the contests of National Plan for Science Degree for physics. Despite the total commitment required to all (students, teachers, professors) for about 10 hours each day, everyone usually is very satisfied at the end of the school.

During these years, the school network increased from initial 9 schools up to the actual 18, more than 250 students participated, initial 3 expert teachers becomes now about 8 and some of them started few years ago like young teachers.

The Pigelleto's Summer School of Physics plays a central role for orienting students toward physics in our territorial area. The main goal is the excellent feedback that returns from students, their families and teachers. In the last years, the applications are increased up to double the available positions.

At the end of 2011, the Pigelleto's facility needed a roof's repairing and other major renovations. It was not possible to find a suitable structure for hosting the summer school in 2012. Thus, in 2013 the school of physics moved in a magnificent new accommodation: the former episcopal seminary of Pienza.

Training for teachers works very well. Young teachers benefit from discussion on methodologies and design and from direct experience in laboratory with expert teachers. On the other side, we identified the following critical points:

- often teachers are obliged to interrupt the stage for going back to school service;
- young teachers (usually with no permanent position) can be called for a temporary job during the school;
- few teachers in-service participate to the summer school (they usually prefer not to accompany their student);
- more care is needed in order to involve teachers from pre-service Formative Active Training course for teaching qualification.

On the other side, many activities inspired by the school entered in common practice or became the starting point for developing learning paths to be used in the ordinary didactics at school. Especially young teachers are more active in this direction.

In conclusion, the summer school of physics is an excellent action for orienting students, for improving physics education in select topics and it is very effective as informal training for young teachers. Sometimes, summer school's activities inspires teachers for new learning paths to include in school's didactic but it is difficult to obtain a large participation of in-service teachers.

## REFERENCES

Benedetti, R., Mariotti, E., Montalbano, V., & Porri A., (2011), Active and cooperative learning paths in the Pigelleto's Summer School of Physics, in *Proceedings of Twelfth International Symposium Frontiers of Fundamental Physics* (FFP12), Udine 21-23 November 2011, arXiv:1201.1333v1 [physics.ed-ph]

Bonwell, C.C, and J. A. Eison. (1991). Active Learning: Creating Excitement in the Classroom. (*ASHE-ERIC Higher Education Report No. 1*, 1991) Washington, D.C.

Convert, B. (2005) Europe and the Crisis in Scientific Vocations, *European Journal of Education, 40(4),* 361-366.

Cuseo, J. (1992). Cooperative learning vs small group discussions and group projects: the critical differences, *Cooperative Learning and College Teaching*, 2(3).

Czujko, R. (2002), *Enrollments and Faculty in Physics*, pp. 1-11 http://www.aip.org/statistics/trends/reports/june9talk.pdf, accessed 2012 March.

National Science Board, (2007) The Message of the 2004 *S & E Indicators: An Emerging and Critical Problem of the Science and Engineering Labor Force , 1*, http://www.nsf.gov/statistics/nsb0407/, accessed 2012 March.

Montalbano, V., (2012), Fostering Student Enrollment in Basic Sciences: the Case of Southern Tuscany, in *Proceedings of The 3rd International Multi-Conference on Complexity, Informatics and Cybernetics: IMCIC 201*2, ed. N. Callaos et al, 279, arXiv:1206.4261 [physics.ed-ph].

Montalbano, V., Benedetti, R., Mariotti, E., Mariotti, M. A., & Porri, A (2012), Attempts of transforming teacher practice through professional development, *The World Conference on Physics Education* (WCPE), July 1-6, 2012, Istanbul, arXiv:1212.2174 [physics.ed-ph].

Mulvey P. J., & Nicholson, S. (2008). Physics Enrollments: Results from the 2008 Survey of Enrollments and Degrees, *AIP Focus On*, 2011, 1-10, http://eric.ed.gov/PDFS/ED517370.pdf, accessed 2012 March.

Johnson, D.W., & Johnson, R.T. (1999). Making cooperative learning. *Theory into Practice, 38(2)*, 67-73.

Sassi, E., Chiefari, G., Lombardi, S., & Testa, I. (2012), Improving scientific knowledge and laboratory skills of secondary school students: the Italian Plan "Scientific Degrees"*, The World Conference on Physics Education* (WCPE), July 1-6, 2012, Istanbul.

Stefanel, A., Michelini, M., Altamore, A., Bochicchio, M., Bonanno, A., De Ambrosis, A., …Stella, R. (2012), The Italian project IDIFO3 (Innovazione Didattica in Fisica e Orientamento 3), *The World Conference on Physics Education* (WCPE), July 1-6, 2012, Istanbul.